\documentclass[%
 reprint,
 amsmath,amssymb,
 aps,
]{revtex4-2}

\usepackage{graphicx}
\usepackage{dcolumn}
\usepackage{bm}
\usepackage{xfrac}

\usepackage[version=4]{mhchem} 

\counterwithout{figure}{section}

\usepackage{mathtools}
\mathtoolsset{centercolon}  

\usepackage{xcolor}

\renewcommand{\phi}{\varphi}

\newcommand{\abs}[1]{\left \lvert #1 \right \rvert}%

\renewcommand{\rm}{\mathrm}

\newcommand{\bs}[1]{\boldsymbol{#1}}

\newcommand{\etal}{\textit{et al.}}

\begin{document}

\preprint{APS/123-QED}

\title{Temperature Compensation through Kinetic Regulation in Biochemical Oscillators}

\author{Haochen Fu}
\affiliation{Department of Physics, University of California San Diego, 9500 Gilman Dr, La Jolla, CA 92093, USA}
\author{Chenyi Fei}%
\affiliation{Department of Mathematics, Massachusetts Institute of Technology, 182 Memorial Dr, Cambridge, MA 02139, USA}
\author{Qi Ouyang}
\affiliation{The State Key Laboratory for Artificial Microstructures and Mesoscopic Physics, School of Physics, Peking University, Beijing 100871, PRC}
\author{Yuhai Tu}
 \email[Corresponding author: ]{yuhai@us.ibm.com}
\affiliation{IBM T. J. Watson Research Center, Yorktown Heights, NY 10598, USA}





\begin{abstract}
Nearly all circadian clocks maintain a period that is insensitive to temperature changes, a phenomenon known as temperature compensation (TC). Yet, it is unclear whether there is any common feature among different systems that exhibit TC.  From a general timescale invariance, we show that TC relies on existence of certain period-lengthening reactions wherein the period of the system increases strongly with the rates in these reactions. By studying several generic oscillator models, we show that this counter-intuitive dependence is nonetheless a common feature of oscillators in the nonlinear (far-from-onset) regime where the oscillation can be separated into fast and slow phases. The increase of the period with the period-lengthening reaction rates occurs when the amplitude of the slow phase in the oscillation increases with these rates while the progression-speed in the slow phase is controlled by other rates of the system. The positive dependence of the period on the period-lengthening rates balances its inverse dependence on other kinetic rates in the system, which gives rise to robust TC in a wide range of parameters. We demonstrate the existence of such period-lengthening reactions and their relevance for TC in all four model systems we considered. Theoretical results for a model of the Kai system are supported by experimental data. A study of the energy dissipation also shows that better TC performance requires higher energy consumption. Our study unveils a general mechanism by which a biochemical oscillator achieves TC by operating at regimes far from the onset where period-lengthening reactions exist.  
\end{abstract}

\maketitle

\section{Introduction}
Biological systems are subject to temperature changes in their environments. Most biochemical reactions are dependent on temperature $T$, with their kinetic rate constants $k$ obeying the Arrhenius law \cite{Arrhenius1889-um,Phillips2012-hr}:
\begin{equation}
    k(T) = A e^{-\frac{E}{k_\rm{B}T}}, \label{eq:arrhe}
\end{equation}
where $A$ is a temperature-independent prefactor, $k_\rm{B}$ is the Boltzmann constant, and $E$ is the activation energy. The temperature sensitivity of a chemical reaction can be characterized by its $Q_{10}$ factor, which measures the change in the reaction rate when the temperature is increased by 10~K. For typical biochemical reactions, the activation energy is $E\sim 20~k_\rm{B}T_0$ \cite{Milo2015-sw}, leading to a $Q_{10}\approx 2$ under room temperature $T_0\approx 300$~K – the rate of a biochemical reaction approximately doubles upon 10 K increase in temperature (around room temperature). Such temperature dependence occurs in various biological processes, such as bacterial cell growth \cite{Knapp2022-tr} and vertebrate muscle contraction\cite{Bennett1984-gp}. 

Given the large $Q_{10}$ values for individual reaction rates, it is remarkable that nearly all circadian clocks can maintain a relatively constant period over a wide range of temperatures, a phenomenon known as temperature compensation (TC) \cite{Pittendrigh1954-un,Johnson2021-ds,Hogenesch2011-tr}. Specifically, the $Q_{10}$ value of the period of a circadian clock is between 0.9 and 1.1 \cite{Johnson2014-jk}; namely, a 10~K increase in temperature only results in a 10\% change in the period of the circadian clock, which is much smaller than the typical 2-fold changes for individual reaction rates. While some previous studies attributed TC to certain genes or proteins whose activities are insensitive to temperature \cite{Isojimaa2009-ug,Terauchi2007-ol,Murakami2008-eu}, biological clocks are not simply insensitive to temperature, e.g., it is well known that the clock can be readily entrained by external periodic temperature signals \cite{Zimmerman1968-vg,Winfree1972-tv,Yoshida2009-fm}.

In general, dynamics of biochemical oscillations are governed by biochemical reaction networks with multiple bio-molecules interacting through different biochemical reactions. As a result, the period $P({\bs k})$ of the oscillation should be a function of all the reaction rates ${\bs k}=\{k_i\}$ with $i(=1,2,...)$ labeling the individual reactions.  Here, we define the “period sensitivities”
\begin{equation}
    C_i\equiv\frac{\partial\ln P}{\partial\ln k_i}\bigg|_{T=T_0}, \label{eq:C_i}
\end{equation}
to describe the dependence of the period $P$ on the i$^\mathrm{th}$ reaction rate $k_i$ -- reactions are referred to as period-lengthening ($C_i>0$) or period-shortening reactions ($C_i<0$), respectively. 

From a systems perspective, while each individual reaction rate can be sensitive to temperature with a high $Q_{10}$, TC could in principle arise from a balance between period-lengthening reactions and period-shortening reactions \cite{Ruoff1992-vv,Ruoff2007-gz,Kurosawa2005-fw}. To see this, we can approximate $Q_{10}$ of the period as $Q_{10}= \frac{P(T_0+10\mathrm{K})}{P(T_0)} \simeq 1 + 10~\mathrm{K} \times \frac{\rm{d} P}{\rm{d}T}|_{T=T_0}$, and use the Arrhenius equation~Eq.~\ref{eq:arrhe} to derive
\begin{equation}
	Q_{10} \simeq 
	1+\frac{10~\mathrm{K}}{T_0}\sum_{i}C_i({\bs k})E_i, \label{eq:Q10_taylor}
\end{equation}
where the summation is over all the reactions, $E_i$ is the activation energy of the i$^\mathrm{th}$ reaction in the unit of $k_\rm{B}T_0$. Perfect TC ($Q_{10}=1$) is achieved by having  $\sum_{i}C_iE_i =0$, which we will refer to as the ``perfect balancing scenario" hereafter. Since the activation energies are positive ($E_i>0$), perfect TC requires the cancellation of the weighted contributions $C_iE_i$ from the period-lengthening reactions ($C_i>0$) and the period-shortening reactions ($C_i<0$). 

While this general systems view of achieving TC is appealing, it is unclear how a biochemical system (network) is able to implement the perfect balancing scenario. Previous studies tackled the problem from the perspective of the reaction network structure \cite{Wu2017-vy,Francois2012-gl,Hong2007-as,Ruoff2007-gz}. For example, it was found that oscillators with positive feedback are more likely to achieve TC than those with purely negative feedback \cite{Wu2017-vy,Baum2016-ru}. However, even for a preferred network topology for TC, the performance of TC is still highly dependent on kinetic rates in the network~\cite{Wu2017-vy}. In addition, it is typically easier for organisms to achieve a biological function, such as TC, by adjusting specific kinetic rate parameters within an existing network structure rather than modifying the entire network, because the former only needs small modifications in molecular structures, whereas the latter typically requires additional molecular components in the system.  Therefore, it is important to explore strategies for controlling (regulating) kinetic rate parameters in order to achieve TC.

Indeed, kinetic regulation, by which we mean controlling kinetic rates in a temperature-independent way, in general and in particular changing the prefactors in Eq.~\ref{eq:arrhe} by varying certain molecule concentrations in the system, plays an important role in biochemical oscillators. The non-equilibrium nature of limit-cycle oscillation dynamics dictates that the kinetic rates in the underlying biochemical network break detailed balance~\cite{Qian2007-tw}. Furthermore, the kinetic rates must be increased beyond the onset of oscillation, typically a Hopf bifurcation for limit-cycle oscillations \cite{Kuznetsov1998-td}, in order to generate the oscillatory behaviors. However, these basic requirements do not necessarily result in temperature-compensated oscillations. Thus, conditions for TC remain elusive. 

In this paper, we start our investigation with two simplest nonlinear oscillators --- the Van der Pol model and the Brusselator model—representing two fundamental motifs in two-component oscillatory systems: the activator-inhibitor motif and the substrate-depletion motif, respectively \cite{Tyson2002-rf}. We find that kinetic regulation provides a general mechanism wherein increasing certain period-lengthening reaction rate(s), which is possible only far from the onset, can lead to robust TC within a realistic range of activation energies. The energy cost associated with this mechanism is also studied. Finally, we verify the applicability of this general TC mechanism in two realistic biological clocks.

\section{Results}
\subsection{Achieving temperature compensation by kinetic regulation}

\begin{figure}[!t]
\centering
\includegraphics[width=0.48\textwidth]{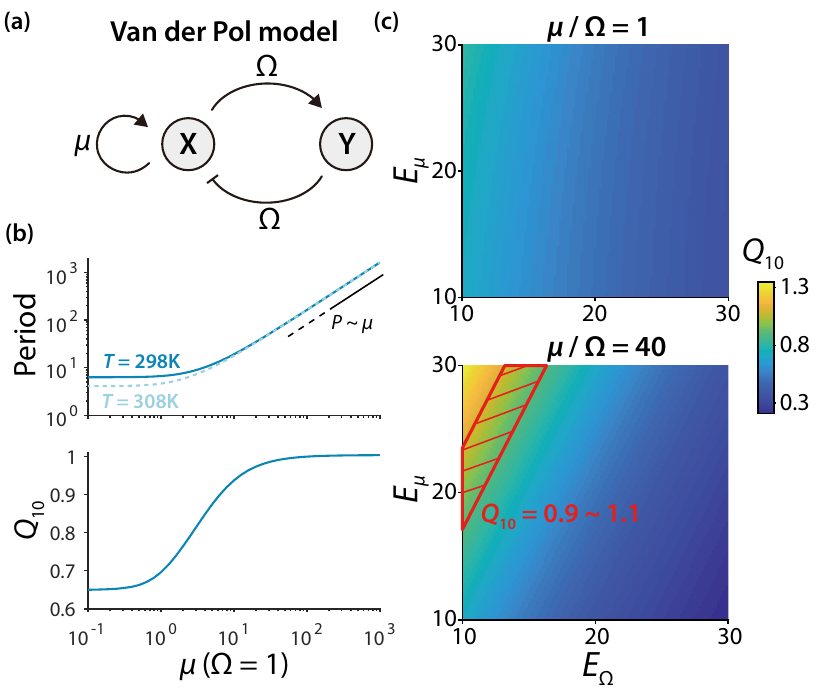}
\caption{\label{fig:1} The Van der Pol (VdP) model can achieve temperature compensation by kinetic regulation. (a) The topology of the VdP model is an activator-inhibitor. (b) The period at the room temperature ($T_0=298$K) and that at 10K above vary with the ``kinetic rate" $\mu$ by varying the prefactor of $\mu$; $Q_{10}$ value is calculated by the ratio of the period at such two temperatures. The activation energies are set at $E_\mu = 26.7$ and $E_\Omega = 13.3$ ($E_\mu \approx 2 E_\Omega$) in the unit of $k_\rm{B} T_0$. (c) The $Q_{10}$ value in the activation-energy space when $\mu \sim \Omega$ and when $\mu \gg \Omega$ (both $\mu$ and $\omega$ are the values at $T=T_0$). The TC regime $0.9\leq Q\leq 1.1$ is highlighted in red. } 
\end{figure}

We start by considering one of the simplest and the most generic nonlinear oscillators — the Van der Pol (VdP) oscillator \cite{Ginoux2012-pq}, described by
\begin{subequations}\label{eq:ODE_VdP}
    \begin{align}
        \frac{\mathrm{d} X}{\mathrm{d} t}&=\mu X(1 - \frac{X^2}{3}) -\Omega Y, \\
        \frac{\mathrm{d} Y}{\mathrm{d} t}&=\Omega X,
    \end{align}
\end{subequations}

which shares a similar network topology as an activator-inhibitor model [Fig.~\ref{fig:1}(a), see also SI Sec.~I(A)] \cite{Tyson2002-rf,Cao2015-xa,Wu2017-vy}. To study the performance of TC in the VdP model, we assume that the rate parameters $\Omega$ and $\mu$ depend on the temperature through the Arrhenius law Eq.~\ref{eq:arrhe} with activation energies $E_\Omega$ and $E_\mu$, respectively. Unless otherwise noted, we study effects of kinetic regulation for TC by varying the temperature-independent prefactors ($A_\mu$, $A_\Omega$, etc.) in Eq.~\ref{eq:arrhe}.    

The VdP model (Eqs.~\ref{eq:ODE_VdP}) undergoes a Hopf bifurcation at $\mu=0$, with a stable limit cycle for $\mu > 0$. As shown in Fig.~\ref{fig:1}(b), the period of the limit-cycle oscillation is sensitive to temperature change when $\mu$ is small or comparable to $\Omega$. However, if $\mu$ becomes much larger than $\Omega$, TC of the period can be achieved, reflected clearly by the $Q_{10}$ value of the period approaching 1 [Fig.~\ref{fig:1}(b)]. Such a realization of TC can also be under the constraint of constant period at $T=T_0$ [see Fig.~S1 in Supplementary Information (SI)].

The good TC performance for large $\mu/\Omega$ as shown in Fig.~\ref{fig:1}(b) is achieved for a particular choice of activation energies. The natural question is how robust is TC with respect to the choices of the activation energy parameters. To address this question, we calculate $Q_{10}$ in the activation-energy space of $E_{\mu}$-$E_{\Omega}$. We limit the range of the activation energies to be between 10 and 30 $k_\rm{B}T_0$, which are relevant for typical biochemical reactions \cite{Milo2015-sw} and also sufficiently large so that the reactions are not trivially temperature-insensitive. When $\mu$ is small or comparable to $\Omega$, $Q_{10}$ is globally small in the activation energy space, and there is no regime that satisfies the good TC requirement $0.9\leq Q_{10}\leq 1.1$, as shown in Fig.~\ref{fig:1}(c). However, when $\mu \gg \Omega$, there exists a band regime where $0.9\leq Q_{10}\leq 1.1$, suggesting that TC can be achieved with a variety of combinations of activation energies [Fig.~\ref{fig:1}(c)]. Indeed, the size of the regime in the activation energy space wherein TC exists ($0.9\leq Q_{10}\leq 1.1$) serves as the most important measure of TC performance for a given kinetic regulation scheme. By this measure, kinetic regulation by  increasing $\mu/\Omega$ improves the TC performance in the VdP model as shown in Fig.~\ref{fig:1}(c). More explicitly, the area in the $(E_{\mu},E_{\Omega})$ space with $0.9\le Q_{10}\le 1.1$ (the shaded area in Fig.~\ref{fig:1}(c)) increases with $\mu/\Omega$ as shown in Fig.~S2 in the SI.

\subsection{The period sensitivity sum rule and the importance of positive period sensitivity for temperature compensation}
\begin{figure}[!t]
\centering
\includegraphics[width=0.48\textwidth]{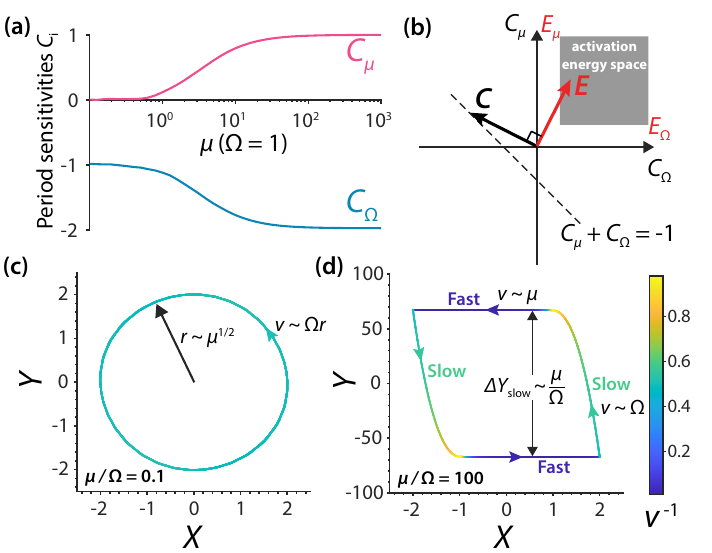}
\caption{\label{fig:2} The TC mechanism in VdP model in the balancing scenario. (a) The period sensitivities vary with $\mu$. (b) The period sensitivity vector $\bs{C}$ is constrained on the $\sum_{i}C_i=-1$ subplane, whereas the activation energy vector $\bs{E}$ is constrained in the activation-energy space. Perfect TC requires $\bs{C}\cdot \bs{E} = 0$. These geometric constraints have made positive period sensitivities to be critical to achieving the balancing scenario and hence TC. (c,d) The amplitude and speed behavior on the limit cycle near the onset ($\mu/\Omega = 0.1$) and far from the onset ($\mu/\Omega = 100$). The color bar shows the inverse of the progression speed, i.e., time per distance, for both panels.} 
\end{figure}

In biochemical reaction systems that exhibit oscillatory behaviors, if all the reaction rates are changed by the same factor $\lambda$, then the period should change by a factor $\lambda^{-1}$. From this time scaling invariance, one can obtain an exact sum rule  among period sensitivities of all the reactions in the system \cite{Ruoff2003-yv,Kurosawa2005-fw}:
\begin{equation}
    \sum_{i}C_i=-1, \label{eq:C_i_sum}
\end{equation}
which serves as a global constraint for all the period sensitivities.
Indeed, the period sensitivities of the VdP model always satisfy $C_\mu + C_\Omega = -1$ [Fig.~\ref{fig:2}(a)]. If we define the period sensitivity vector $\bs{C}$ whose components are the period sensitivities of each reactions, and correspondingly the activation energy vector $\bs{E}$, the sum rule Eq.~\ref{eq:C_i_sum} constrains $\bs{C}$ on a subplane,
whereas the condition for perfect TC requires that $\bs{E}$ is perpendicular to $\bs{C}$, i.e., 
\begin{equation}
    \bs{C}\cdot \bs{E} = 0.\label{eq:C_E_0}
\end{equation}

Given that all the activation energies are positive, TC cannot be achieved in a biochemical network where all components of $\bs{C}$ are negative, which is certainly allowed by the sum rule. Thus, a necessary condition for TC is the existence of period-lengthening reactions with positive period sensitivities, which is a non-trivial requirement in light of the sum rule (Eq.~\ref{eq:C_i_sum}). Quantitatively, if the positive components of $\bs{C}$ are small, the corresponding components of vector $\bs{E}$ must be extremely large compared to other components, and $\bs{E}$ is restricted on the margin of the non-trivial activation-energy space, as illustrated in Fig.~\ref{fig:2}(b) (for a system consisting of two kinetic rates). On the other hand, a large positive component of $\bs{C}$ can move $\bs{E}$ toward the center of the activation-energy space, allowing TC to be achieved by more combinations of activation energies that are physiologically plausible [shaded regime in Fig.~\ref{fig:2}(b)]. Therefore, TC can be achieved more robustly in the presence of large positive period sensitivities.

The general analysis based on the period sensitivity sum rule can be used to understand why TC is systematically improved by regulation of ``kinetic rates" $\mu$ and $\Omega$. As shown in Fig.~\ref{fig:2}(a), when $\mu \lesssim \Omega $, $C_\mu \approx 0$ and $C_\Omega \approx -1$ [see also SI Sec.~I(A) for theoretical derivation]. Hence, the negative $C_\Omega E_\Omega$ cannot be balanced by a nearly zero $C_\mu E_\mu$, leading to a $Q_{10}$ substantially smaller than 1. As $\mu$ increases to $\mu \gg \Omega $, $C_\mu$ increases to 1 while $C_\Omega$ decreases to -2 [see Sec.~I(A) in the SI]. In this case, the balance between $C_\Omega E_\Omega$ and $C_\mu E_\mu$ becomes possible. Especially, to achieve perfect TC, we only need $E_\mu = 2 E_\Omega$, so that $C_\Omega E_\Omega + C_\mu E_\mu = 0$, as is the case in Fig.~\ref{fig:1}(b).

\subsection{Large positive period sensitivity arises from ``oscillation phase separation"}
How does $C_\mu$ become large (close to 1) in the $\mu\gg\Omega$ regime for the VdP oscillator? To address this question, we examine the limit cycle trajectory when $\mu \gg \Omega$. As shown in Fig.~\ref{fig:1}(d), the limit cycle is separated into fast and slow phases (the inverse progression speed, i.e., the time per distance traveled, is shown by the color along the trajectory) and the period is mostly determined by the time spent on the slow phases along the two segments of the $X$-nullcline. The amplitude of the slow phases along the $X$-nullcline, denoted by $\Delta Y_\mathrm{slow}$, is proportional to $\mu/\Omega$ [see Sec.~I(A) in SI for derivation]. \textcolor{black}{Surprisingly, unlike the situation near the onset,} the progression speed in the slow phases is given by $\mathrm{d} Y/\mathrm{d} t=\Omega X \sim\Omega$, insensitive to the change of $\mu$. Indeed, increasing $\mu$ accelerates the progression speed linearly only in the fast phases where the time is negligible [Sec.~I(A) in SI]. Consequently, the period of the oscillation $P\approx 2\Delta Y_\mathrm{slow} / |\mathrm{d}Y/\mathrm{d}t| \sim \mu / \Omega^2$ increases linearly with $\mu$, leading to a large positive $C_\mu$. 

\begin{figure}[!t]
\centering
\includegraphics[width=0.48\textwidth]{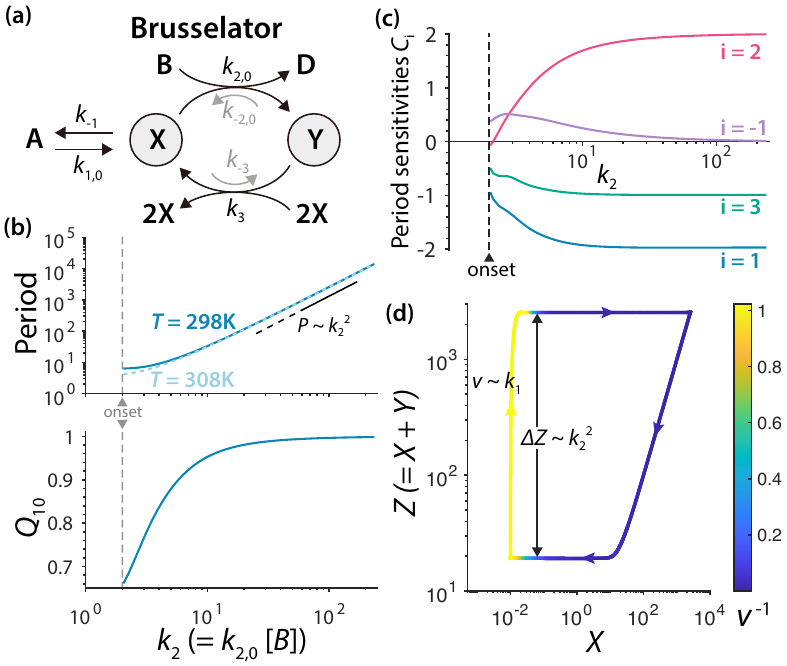}
\caption{\label{fig:3} TC can be achieved by kinetic regulation in Brusselator. (a) The reversible Brusselator model. The concentrations of A, B, and D are kept constant over time. When the two reverse reaction rates, $k_{-2} = k_{-2,0} [D]$ and $k_{-3}$, are negligible, the Brusselator becomes its original irreversible version. (b) In the irreversible Brusselator, the period at two temperatures (298K and 308K) and correspondingly $Q_{10}$ value of the period vary with the kinetic rate $k_2$. The activation energies are set as $E_1 = 15$, $E_2 = 25$, and $E_3 = E_{-1} = 20$ ($k_\rm{B}T_0$). (c) In the irreversible Brusselator, the enhancement of TC is caused by the emergence of large positive period sensitivity $C_2$, which increases from nearly 0 to 2 with the increase of $k_2$. (d) The limit cycle of the irreversible Brusselator far from the onset with the color bar showing the inverse of the progression speed. The large positive period sensitivity of $k_2$ is a result of a $k_2$-sensitive amplitude and a $k_2$-insensitive progression speed of the slow phase that dominates the period.
}
\end{figure}

Overall, the separation of the fast and slow phases along the limit cycle trajectory in the parameter regime far beyond the oscillation onset, a phenomenon which we call ``oscillation phase separation" (OPS), plays an important role in creating a large positive period sensitivity that is crucial for TC. In the VdP model, the emergence of the slow phase(s) with a $\mu$-insensitive progression speed and a $\mu$-increasing amplitude is the origin of a period-lengthening rate with a large positive period sensitivity. In general, OPS or equivalently the separation of timescales is a common feature in a large class of oscillatory systems known as relaxation oscillators~\cite{strogatz1994nonlinear,wang1999relaxation} exemplified by the VdP model. In the rest of the paper, we show that this general mechanism for TC based on OPS applies for other more realistic biochemical oscillatory systems. 

\subsection{Temperature compensation occurs in Brusselator through ``oscillation phase separation"}
While the VdP model studied in previous sections represents the activator-inhibitor motif, in this section, we study TC in another class of oscillators with the substrate-depletion motif~\cite{Tyson2002-rf,Cao2015-xa,Wu2017-vy} as represented by the reversible Brusselator model \cite{Nicolis1977-dr,Qian2002-rn,Fei2018-rb}. This classical model describes the nonlinear autocatalytic reactions [Fig.~\ref{fig:3}(a)] 
\[
\ce{A <=>[\ce{$k_{1,0}$}][\ce{$k_{-1}$}] X},~ \ce{X + B <=>[{$k_{2,0}$}][{$k_{-2,0}$}] Y + D},~ \ce{2X + Y <=>[{$k_3$}][{$k_{-3}$}] 3X},
\]
where the forward and reverse kinetic rates depend on temperature via Eq.~\ref{eq:arrhe}, and A, B, and D have constant concentrations $[A]$, $[B]$, and $[D]$, respectively. 

For simplicity, we first neglect the reverse kinetic rates $k_{-2,0}$ and $k_{-3}$, and the model becomes the original irreversible Brusselator \cite{Nicolis1977-dr,Qian2002-rn}. Sustained oscillation occurs when the kinetic rate $k_2 = k_{2,0}[B]$ exceeds the Hopf bifurcation critical point $k_2^c$, beyond which the amplitude of limit-cycle oscillation further increases with $k_2$. Remarkably, increasing $k_2$ can also enhance the performance of TC [Fig.~\ref{fig:3}(b)], and the good TC performance at large $k_2$ is robust as it can be realized in a wide range of activation energies (see Fig.~S4 in SI for details). In the Brusselator model, the period sensitivity $C_2$ for reaction $k_2$ is positive and it increases from nearly 0 to around 2 with increasing $k_2$[Fig.~\ref{fig:3}(c)], which leads to good TC performance at larger values of $k_2$. 

To understand the positive period sensitivity of $k_2$ when $k_2 \gg k_2^c$, we analyze the limit cycle trajectory in the $X$-$Z$ phase space where $Z=X+Y$ denotes the total concentration of $X$ and $Y$. Similar to the VdP model, the limit-cycle trajectory also exhibits the oscillation phase separation behavior at large values of $k_2$ with a slow phase and three relatively fast phases as shown in Fig.~\ref{fig:3}(d). During the slow phase, $X$ is kept at a roughly constant but low level $X\ll k_{1,0}[A]/k_{-1}$, resulting in a nearly constant progression speed $v_\rm{slow} \approx \mathrm{d} Z/\mathrm{d} t = k_{1,0}[A] - k_{-1} X \approx k_{1,0}[A]$, independent of $k_2$ [see Sect. I(B) in SI for details]. On the other hand, the amplitude of the slow branch $\Delta Z_\rm{slow}$, and hence the time on the slow phase $\tau_\rm{slow}\approx \Delta Z_\rm{slow}/v_\rm{slow}$, asymptotically increases with $k_2$ as $k_2^2$ [see Sec.~I(B) in SI for detailed derivations]. Indeed, the time in the other phases negatively depends on $k_2$, but this period-shortening effect is negligible as the times spent in these fast phases are much shorter than that in the slowest phase [see Sec.~I(B) in SI for detailed derivations]. As a result, the period can be approximated by $\tau_\rm{slow}$ that scales with $k_2^2$, leading to the period sensitivity $C_2\approx 2$ when $k_2 \gg k_2^c$. 

\subsection{Energetic cost of temperature compensation via kinetic regulation}
\begin{figure}[!t]
\centering
\includegraphics[width=0.48\textwidth]{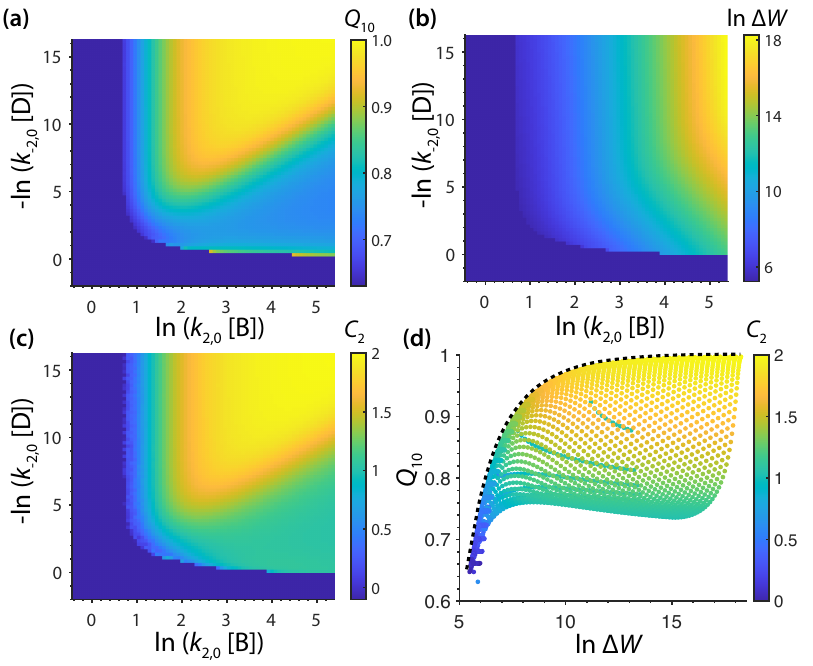}
\caption{\label{fig:4} In the reversible Brusselator, kinetic regulation consumes free energy to enhance the TC performance. (a,b,c) $Q_{10}$, energy dissipation $\Delta W$, and period sensitivity $C_2$ in the parameter space spanned by $k_2 = k_{2,0}[B]$ and $k_{-2} = k_{-2,0} [D]$ (log scale). The dark blue regime does not have sustained oscillation. Activation energies are set as $E_1 = 15$, $E_2 = 25$, and $E_3 = E_{-1} = E_{-2} = E_{-3} = 20$ ($k_\rm{B}T_0$). (d) Scatter plots of $Q_{10}$ against the energy cost $\ln \Delta W$ according to (a) and (b), with the color showing the period sensitivity $C_2$ of (c). The dashed curve shows the upper bound of the scattered plot.} 
\end{figure}

In a biochemical system, kinetic regulation is often achieved by regulating rate-associated molecular concentrations. For example, in the Brusselator model, regulation of $k_2$ can be easily implemented by changing the concentration $[B]$. However, it consumes free energy to maintain a molecular concentration at a nonequilibrium level \cite{Qian2007-tw}. Thus, kinetic regulations of reaction rates incurs an energy cost. To study the energy cost of TC enhancement in Brusselator, we consider the full reaction network with finite reverse reaction rates, $k_{-2} = k_{-2,0} [D]$ and $k_{-3}$. The chemical potential difference between B and D (in units of $k_\mathrm{B}T_0$) reads
\begin{equation}
    \Delta \mu_\rm{DB} = \ln \frac{k_{2,0}[B]k_3}{k_{-2,0}[D]k_{-3}},
\end{equation}
which characterizes the irreversibility of the system \cite{Qian2007-tw,Fei2018-rb} -- when $\Delta \mu_\rm{DB}=0$, the system is at equilibrium, and sustained oscillation cannot be achieved. Increasing $\Delta \mu_\rm{DB}$ drives the system away from equilibrium and enables oscillation, which can be achieved by either increasing $[B]$ or decreasing $[D]$. To quantify the energy cost, we compute the free-energy dissipation per oscillation period \cite{Qian2007-tw,Fei2018-rb}, $\Delta W$, given by
\begin{equation}
	\Delta W =\int_{0}^P\sum_{i}[J_i^+(t)-J_i^-(t)]\ln\frac{J_i^+(t)}{J_i^-(t)}\mathrm{d}t, \label{eq:Delta_W}
\end{equation}
where $J_i^+(t)$ and $J_i^-(t)$ denote, respectively, the forward and the reverse fluxes of the i$^\rm{th}$ reversible reaction pair at time $t$. 

For the reversible Brusselator model, we compute the $Q_{10}$ of the period, the period sensitivity $C_2$ of $k_2$, and the energy dissipation $\Delta W$ for varying reaction rates $k_2$ and $k_{-2}$, as shown in Figs. \ref{fig:4}(a,b,c). Remarkably, either increasing $[B]$ or decreasing $[D]$ can increase the positive period sensitivity $C_2$, and consequently enhance TC [Figs. \ref{fig:4}(a, c)]. A perfect TC is achieved in the limit of extremely large $[B]$ and extremely small $[D]$ where the system is far from equilibrium. On the other hand, the free-energy dissipation per cycle ($\Delta W$) is also increased by $[B]$ and decreased by $[D]$ [Fig.~\ref{fig:4}(b)]. If we plot $Q_{10}$ against $\Delta W$ [Fig.~\ref{fig:4}(d)] for different choices of $[B]$ and $[D]$, we find that the upper bound of $Q_{10}$ is increased by the energy cost. Our results indicate that kinetic regulation for TC costs energy and biochemical oscillators can exploit free energy dissipation to enhance TC performance by kinetic regulation.

\subsection{Kinetic regulation and ``oscillation phase separation" for temperature compensation in circadian clocks}
\begin{figure*}[!t]
\centering
\includegraphics[width=0.75\textwidth]{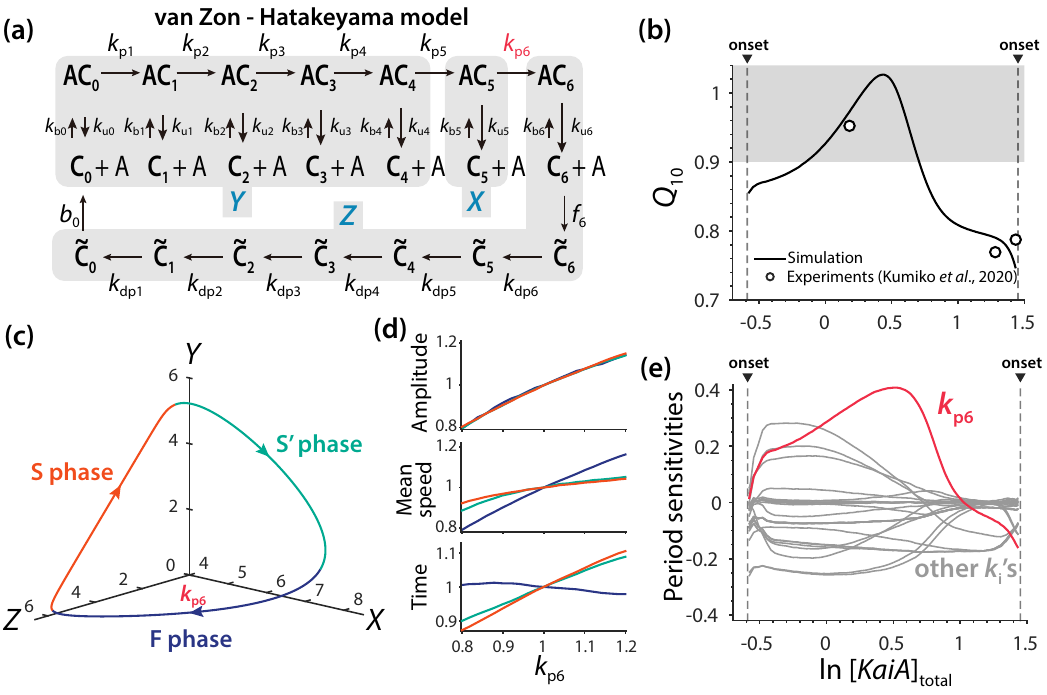}
\caption{\label{fig:5} TC is achieved by regulation of total KaiA concentration in a model of the Kai system. (a) The van Zon-Hatakeyama (vZH) model. KaiC is a hexamer that can be in the active form or the inactive form. During the active state, KaiC can be phosphorylated when binding to KaiA. The binding affinity of KaiA and KaiC decreases with the phosphorylation level of KaiC \cite{Hatakeyama2012-pq}. Based on the oscillation dynamics, different states of KaiC can be grouped into three variables: $X =[C_5]+[AC_5]$, $Y = \sum_{i=0}^4 [C_i]+[AC_i]$, and $Z = [KaiC]_\rm{total} - X - Y$, such that $k_\rm{p6}$ of our interest is explicit in the simplified network of X, Y and Z. (b) $Q_{10}$ of the period varies with the total KaiA concentration. The dots are experimental data from Kumiko \etal~(2020), and the curve is from the simulation of the model with parameters tuned to fit the data [see SI Sec.~I(D) for details]. Activation energies are set to $E_\rm{p6} = 26$, and the other $E_i = 6$ ($k_\rm{B}T_0$). (c) At an intermediate level of total $[KaiA]$ ($\ln [KaiA] = 0.4$), the limit-cycle trajectory projected on the phase subspace of $X$, $Y$, and $Z$ shows oscillation phase separation: S, S' and F. Each phase shows the transition between two variables while the third variable is roughly constant. (d) The relative changes of the amplitude (length of the trajectory), the mean progression speed, and the time on each of the three phases are calculated when $k_\rm{p6}$ is perturbed by $\pm$ 20\%. (e) The period sensitivities of all reaction rates vary with the total [$KaiA$]. Among them, the period sensitivity of $k_\rm{p6}$ (the kinetic rate of the sixth phosphorylation step) has a relatively high positive sensitivity in the middle range of total [$KaiA$]. 
}
\end{figure*}

To study whether the above mechanism of kinetic regulation can be applied to understanding TC in realistic biological systems, we focus on two well-known systems that exhibit circadian rhythms: the \textit{Drosophila} circadian clock and the Kai system in cyanobacteria circadian clocks. The former represents the family of trascription-translational circadian clocks, while the latter stands for the post-transcriptional circadian clocks. Both systems have been reported to have TC \cite{Pittendrigh1954-un,Rothenfluh2000-bk,Williams2001-ue,Kondo1993-lw,Nakajima2005-jb,Tomita2005-bl}. 

We first investigate the model proposed by Tyson \etal~\cite{Tyson1999-lr,Tyson2002-rf} for the \textit{Drosophila} circadian clock, which has been reduced to a 2-species model consisting of only the \textit{per} mRNA and total PER proteins [see SI Sec.~I(C)]. The architecture of the Tyson model contains a substrate-depletion motif [see SI Sec.~I(C)] the same as the Brusselator model studied in the previous subsections. Similar to the Brusselator model, we showed that increasing one of the kinetic rates, specifically the phosphorylation rate of PER monomer $k_1$, drives the system away from the onset and the period sensitivity to $k_1$ has a large positive value at large $k_1$ [see Sec.~I(C) and Fig.~S5 in the SI for details]. As a result, TC can be achieved through the same general OPS mechanism by kinetic regulation of $k_1$.

To confirm that the OPS mechanism of TC also works in more complex high-dimensional systems, we next focus on the van Zon-Hatakeyama (vZH) model for the Kai system \cite{Van_Zon2007-ft,Hatakeyama2012-pq}. In this model, the timekeeper protein KaiC forms a hexamer that can switch between active and inactive conformations (states) [Fig.~\ref{fig:5}(a)]. Active KaiC has seven phosphorylation states corresponding to the number of phosphorylated monomers from 0 to 6. Each phosphorylation step requires an active KaiC to bind to the enzyme KaiA, with the binding affinity decreasing as the phosphorylation level of KaiC increases, which is crucial for synchronization of individual KaiC hexamers. On the other hand, inactive KaiC can spontaneously dephosphorylate. The transitions between active and inactive states occur only in the fully phosphorylated state (between $C_6$ and $\tilde{C}_6$) or in the fully dephosphorylated state (between $\Tilde{C}_0$ and $C_0$). 

Previously, Hatakeyama \& Kaneko investigated the TC property in this model \cite{Hatakeyama2012-pq}. Their results suggested that achieving TC relies on the competition for limited KaiA among various active KaiC phosphorylation states. However, their results also showed that this limited-KaiA scheme for TC is effective only in specific parameter regime \cite{Hatakeyama2012-pq}. To understand this observation, we hypothesize that the Kai system can achieve TC through kinetic regulation of total KaiA concentrations, since KaiA is involved in all phosphorylation rates. Indeed, recent experiments have shown that TC is achieved at a moderate level of KaiA, but not near the onset of oscillation at a high KaiA level \cite{Ito-Miwa2020-kk}. Another {\it in vivo} study demonstrates that the TC property is abolished in a mutant lacking active KaiA \cite{Kawamoto2020-ti}. These findings indicate that TC exists only within an intermediate range of KaiA levels. 

To better understand the dependence of TC on the concentration ($[KaiA]$), we performed simulations based on the vZH model [see Sec.~I(D) in SI for model details]. As we vary the KaiA concentration, there are two onsets of oscillation: the lower onset is determined by the minimum amount of KaiA needed for phosphorylation reactions; and the higher onset is determined by the maximum KaiA concentration beyond which synchronization fails due to lack of competition among individual KaiC hexamers for KaiA. We computed $Q_{10}$ for the oscillation period in the range of KaiA concentration between these two onsets. Consistent with the experimental data \cite{Ito-Miwa2020-kk,Kawamoto2020-ti}, we found that the vZH model fails to achieve TC ($Q_{10}< 0.9$) when $[KaiA]$ is close to the two onsets, however, the value of $Q_{10}$ peaks around $1$ (perfect TC) at intermediate levels of KaiA away from the two onsets as shown in Fig.~\ref{fig:5}(b). Similar to previous models considered in this paper, the non-monotonic change in $Q_{10}$ with increasing KaiA levels relies on the crucial kinetic rate $k_\rm{p6}$ of the sixth phosphorylation step, whose period sensitivity is negative or close to zero near the onsets of oscillation but becomes the largest positive period sensitivity in the intermediate range of $[KaiA]$ [Fig.~\ref{fig:5}(e)]. 

To investigate how the period sensitivity of $k_\rm{p6}$ becomes large and positive in the intermediate range of $[KaiA]$, we combine the state variables into three groups: $X =[C_5]+[AC_5]$, $Y = \sum_{i=0}^4 [C_i]+[AC_i]$, and $Z = [KaiC]_\rm{total} - X - Y$ [Fig.~\ref{fig:5}(a)], and study the limit cycle projected in the three dimensional phase space [Fig.~\ref{fig:5}(c)]. The oscillation is evidently separated into three phases: S, S', and F. Each phase represents the conversion between two species with the third species roughly constant. For example, the S phase is when $Z$ is converted to $Y$ while $X$ remains unchanged. We investigate how the amplitude (i.e., the trajectory length), mean progression speed, and the time of each phase vary with $k_\rm{p6}$. As shown in Fig.~\ref{fig:5}(d), the amplitudes of all three phases increase with $k_\rm{p6}$ in a similar behavior; as discussed previously, this amplitude-increasing effect is the necessary condition for a period-lengthening reaction. What distinguishes the three phases is the behavior of the mean progression speeds. While the mean progression speed of the F phase scales with $k_\rm{p6}$ in a similar behavior to the amplitude, those of the S phase and the S' phase are relatively insensitive to changes in $k_\rm{p6}$. As a result, the time in the F phase is insensitive to changes in $k_\rm{p6}$, while the time in S and S' phases increases with $k_\rm{p6}$. Therefore, OPS is the key to generating a relatively large positive sensitivity for $k_\rm{p6}$: $k_\rm{p6}$ only accelerates the F phase, a small part of the period, so that this speeding effect of $k_\rm{p6}$ cannot fully counterbalance its amplitude increasing effect. We also observe that this OPS mechanism is absent near the onset of oscillation. In this regime, an increase in $k_\rm{p6}$ simultaneously elevates both the amplitude and the progression speed of the entire limit cycle, leading to a small period sensitivity (see Fig.~S6 in the SI for details).

\section{Discussion}
It has been a long-standing problem whether circadian clocks in various organisms share certain common mechanism for achieving TC~\cite{Kurosawa2005-fw,Hatakeyama2012-pq,Hatakeyama2015-ux,Kidd2015-ut,Kurosawa2017-zh,Gibo2019-lu,Kon2021-ga}. 
In this paper, we provide a novel perspective to tackle this question. By studying four models of biochemical oscillations, we have demonstrated that kinetic regulation can serve as a general mechanism for achieving TC. We focus on cases where kinetic regulation can be implemented by regulation of molecular concentrations, such as $[B]$ or $[D]$ in the Brusselator model and total $[KaiA]$ in the vZH model. The underlying mechanism of such kinetic regulation is the emergence of large positive period sensitivities. A large positive period sensitivity of a particular rate $k$ is non-trivial given the general period sensitivity sum rule that constraints the sum of all period sensitivities to be $-1$. Here, we show that a large positive period sensitivity can be caused by the ``oscillation phase separation" (OPS) phenomenon wherein the whole oscillation trajectory is separated into different phases in time and the dominant slow phase(s) has an amplitude that increases with a particular rate $k$ and a progression speed that is independent of $k$.

To achieve ``oscillation phase separation" (OPS) that is critical for TC, a necessary condition is that the system must be driven far from the onset of limit-cycle oscillation (Hopf bifurcation). The reason is that near Hopf bifurcation the oscillation is sinusoidal \cite{Kuznetsov1998-td}, leading to a nearly uniform progression speed along the whole limit cycle without OPS. In general, kinetic regulation, i.e., controlling certain key kinetic rates can lead to TC by driving the system deep into the oscillatory regime away from the onset where the OPS phenomenon exists. Even though we can not prove it rigorously, OPS always occurs in the nonlinear regime of all the different biochemical oscillators we studied. 

Naturally, driving a system into its nonlinear regime far from the onset requires extra free energy dissipation, as we show explicitly in the case of the Brusselator model where higher free energy dissipation is needed to achieve better TC performance [see Fig.~4(d)]. In the Kai system, sustained oscillation consumes ATP \cite{Nakajima2005-jb}. However, most experiments for TC study in the KaiC system maintain a sufficiently high level of ATP \cite{Tomita2005-bl,Murayama2017-kw,Ito-Miwa2020-kk}, which makes the effect of free energy cost for TC hard to assess. In another study, the Kai system is shown to have metabolic compensation \cite{Phong2013-dz}; that is, the period is insensitive to the change in ATP level. This raises the question of why the cell maintains a high ATP level and consumes more energy for a circadian clock than the minimum energy required for the onset of oscillation. We speculate that kinetic regulation of the ATP/ADP ratio may be used to drive the system away from the onset where the ``oscillation phase separation" mechanism for TC is at play. If the ATP/ADP ratio is lowered to a level close to the onset, the Kai     system could lose its TC property. This prediction can be tested in future experiments involving the Kai system and possibly other circadian clocks.

Most biological oscillators contains a small number of molecules and they need to operate in room temperature environment with large thermal fluctuations. To suppress noise caused by the stochastic biochemical reactions, biological oscillators deploy various control mechanisms in order to carry out their functions, e.g., accurate oscillation, high entrainability to external signals, and synchronization \cite{Cao2015-xa,Fei2018-rb,Zhang2020-lh}, all of which require kinetic rate regulation and extra free energy dissipation to drive the system beyond the onset of oscillation. Our present work shows that TC, the robustness of the oscillation period against temperature fluctuation, another important function for a class of biological oscillators, can also be achieved by kinetic regulation with the corresponding energy dissipation. These results suggest that the energy-assisted kinetic regulation scheme may serve as a general mechanism for biological networks in particular biochemical oscillators to control internal and external fluctuations to enhance their functional performance. \textcolor{black}{This not only deepens our understanding of biological clocks but also offers valuable insights into constructing synthetic clocks with relevant biological functions.}

\bibliography{TC_manuscript}

\newpage
\onecolumngrid
\begin{center}
\textbf{\large Supplementary Information}
\end{center}
\setcounter{equation}{0}
\setcounter{figure}{0}
\setcounter{section}{0}
\renewcommand{\theequation}{S\arabic{equation}}
\renewcommand{\thefigure}{S\arabic{figure}} 
\renewcommand{\thesection}{\Roman{section}}
\renewcommand{\thesubsection}{\Alph{subsection}}

\section{Supplementary Notes}

\subsection{Derivation of Period Sensitivities in the VdP Model}

The original Van der Pol model is written as the following 2nd-order ordinary differential equation (ODE):
\begin{equation}
	\frac{\mathrm{d}^2 X}{\mathrm{d} t^2}+\mu(X^2-1)\frac{\mathrm{d} X}{\mathrm{d} t}+\Omega^2X=0,\label{eq:VdP_2nd_order}
\end{equation}
where $\mu$ and $\Omega$ are two kinetic rate parameters that have the unit of $[t^{-1}]$. Using the Lienard representation \cite{strogatz1994nonlinear}, by defining $\frac{\mathrm{d} Y}{\mathrm{d} t} = \Omega X$, we obtain the two-dimensional 1st-order ODE Eqs.~(4) in the main text. 

A natural order parameter of this system is $\nu\equiv\frac{\mu}{\Omega}$. To facilitate analysis, we now calculate the period sensitivities $C$ (as defined in the main text) of $\mu$ and $\Omega$ in the two limits: (1) $\nu\ll 1$, and (2) $\nu \gg 1$.

First, when $\nu\ll 1$, the VdP model is close to a harmonic oscillator. Thus, we can expand the angular velocity ($\omega\equiv 2\pi/P$) around the natural frequency $\Omega$ in a series of $\nu$:
\begin{equation}
    \omega = \Omega [1 + a_1 \nu + a_2 \nu^2 + O(\nu^3)]. \label{eq:omega_expand}
\end{equation}
To determine the leading terms, we notice that the system is symmetric under the transformation: $X^* =-X, t^* = -t, \mu^* = -\mu$ ($\nu^* = -\nu$). Under this transformation, the new angular velocity should have the same form as Eq.~\ref{eq:omega_expand},
\begin{equation}
    \begin{split}
        \omega^* &= \Omega [1+ a_1 \nu^* + a_2 \nu^{*2} + O(\nu^{*3})] \\
        &= \Omega [1 - a_1 \nu + a_2 \nu^2 + O(\nu^3)], \label{eq:omega_star}
    \end{split}
\end{equation}
Since the systems are identical under this transformation, we have $\omega^* = \omega$. Thus, all the odd terms of $\nu$ in Eq.~(\ref{eq:omega_expand}) must vanish, leading to
\begin{equation}
    \omega = \Omega [1+ a_2 \nu^2 + O(\nu^4)]. \label{eq:omega_expand_2}
\end{equation}
Indeed, using perturbation theory \cite{Verhulst2006-fu}, we obtain that $a_2 = -1/16$.

Therefore, based on the definition of the period sensitivities, we have 
\begin{equation}
    \begin{split}
        C_\mu &= -\frac{\partial\ln \omega}{\partial \ln \mu}= \frac{1}{8}\nu^2 + O(\nu^4),\\
        C_\Omega &= -\frac{\partial\ln \omega}{\partial \ln\Omega} = -1 - \frac{1}{8}\nu^2 + O(\nu^4).
    \end{split}
\end{equation}
Thus, $C_\mu$ is 0 when $\nu=0$ and gradually increases with $\nu$, consistent with our simulation results in main Fig.~2(a).

Second, when $\nu\gg 1$ or equivalently $\mu\gg\Omega$, $\mu$ determines the time scale of $\frac{\mathrm{d} X}{\mathrm{d} t}$, while $\Omega$ always determines the time scale of $\frac{\mathrm{d} Y}{\mathrm{d} t}$. Due to this time-scale separation, the limit-cycle oribit is separated into four branches [Fig.~\ref{fig:S3}(a)]: two ``fast branches" where $\frac{\mathrm{d} Y}{\mathrm{d} t}\approx 0$ and the progression speed is large, and two ``slow branches" along the $X$-nullcline where $\frac{\mathrm{d} X}{\mathrm{d} t}\approx 0$ and the progression speed is small.

We now determine the scaling laws of the amplitude and the progression speed on the fast or slow branches. The nullcline of $\frac{\mathrm{d} X}{\mathrm{d} t}= 0$ is given by
\begin{equation}
    Y=\nu X(1-\frac{X^2}{3}),\label{eq:X_null}
\end{equation}
whose extrema $X^*_{\pm}=\pm1, Y^*_{\pm}=\pm \frac{2}{3}\nu$ are the turning points between the fast and slow branches. Thus, the amplitude of the slow branches scales as
\begin{equation}
    \Delta Y_\mathrm{slow} \approx Y^*_+ - Y^*_- =  \frac{4}{3}\nu \sim \nu^1,
\end{equation}
while the amplitude of the fast branches scales as
\begin{equation}
    \Delta X_\mathrm{fast} \approx X(Y^*_{-}) - X^*_{-} = 3 \sim \nu^0.
\end{equation}
On the other hand, the progression speed on the slow branches scales as
\begin{equation}
    v_\rm{slow} \approx \frac{\mathrm{d} Y}{\mathrm{d} t}\bigg|_{1\leq X\leq 2}=\Omega X|_{1\leq X\leq 2} \sim \Omega,
\end{equation}
whereas that on the fast branches reads
\begin{equation}
    v_\rm{fast} \approx \frac{\mathrm{d} X}{\mathrm{d} t}\bigg|_{Y = \frac{2}{3}\nu} =\mu \left[X(1 - \frac{X^2}{3}) -\frac{2}{3}\right] \sim \Omega \nu.
\end{equation}
Thus, the time $\tau$ spent on each branch scales as
\begin{equation}
    \tau_\rm{slow} \sim \frac{\Delta Y_\rm{slow}}{v_\rm{slow}} \sim \frac{\mu}{\Omega^2},\quad \tau_\rm{fast} \sim \frac{\Delta X_\rm{fast}}{v_\rm{fast}} \sim \frac{1}{\mu}.
\end{equation}
Note that $\mu$ lengthens $\tau_\rm{slow}$ but shortens $\tau_\rm{fast}$. Since $\tau_\rm{slow}/\tau_\rm{fast} \sim \nu^2\gg 1$, the slow branches dominate the period. Thus, the acceleration effect of $\mu$ on the fast branches is negligible, given the minimal contribution of $\tau_\rm{slow}$ to the whole period.

Indeed, we can approximate the period by \cite{strogatz1994nonlinear}
\begin{equation}
     \begin{split}
         P &\approx 2\tau_\mathrm{slow}\\
         &= 2\int_{-\frac{2}{3}\nu}^{\frac{2}{3}\nu}\frac{\mathrm{d}Y}{(\mathrm{d}Y/\mathrm{d}t)_\rm{nullcline}}\\
        &=2\int_{2}^{1}\frac{\mathrm{d}Y}{\mathrm{d}X}\bigg{|}_{\rm{nullcline}}\frac{\mathrm{d}X}{\Omega X}\\
        &=2\int_{1}^{2}\frac{\nu(X^2-1)}{\Omega X}\mathrm{d}X\\
        &=\frac{\mu}{\Omega^2}(3-2\ln 2),
     \end{split}
\end{equation}
and subsequently obtain $C_\mu \approx 1$, and $C_\Omega \approx -2$, consistent with main Fig.~2(a).

\subsection{Derivation of Period Sensitivities in the Brusselator}

According to Fig.~3(a) in the main text, the ODEs describing the reversible Brusselator model read
\small{
\begin{subequations}{\label{eq:ODE_bruss}}
    \begin{align}
        \frac{\mathrm{d} X}{\mathrm{d} t}&=k_1-k_{-1}X-k_2X+k_{-2}Y+k_3X^2Y-k_{-3}X^3,\\
        \frac{\mathrm{d} Y}{\mathrm{d} t}&=k_2X-k_{-2}Y-k_3X^2Y+k_{-3}X^3,
    \end{align}
\end{subequations}}where $k_1 = k_{1,0}[A]$, $k_2 = k_{2,0}[B]$, $k_{-2} = k_{-2,0}[D]$, described by the main text. In this section, we will compute the period sensitivities in two limits: (1) near the oscillation onset (Hopf bifurcation) and (2) far from the oscillation onset.

\subsubsection{Near the oscillation onset}

Based on Eqs. (\ref{eq:ODE_bruss}), the fixed point of the system is $X_0=\frac{k_1}{k_{-1}}$ and $Y_0=\frac{k_2+k_{-3}X_0^2}{k_{-2}+k_3X_0^2}X_0$. Near the fixed point, the ODEs can be approximately written as
\begin{equation}
    \frac{\mathrm{d}}{\mathrm{d} t}\begin{pmatrix} x\\y\end{pmatrix}=J\begin{pmatrix} x\\y\end{pmatrix}+O(x^2,y^2),
\end{equation}
where $x=X-X_0$, $y=Y-Y_0$, and $J$ is the Jacobian matrix at the fixed point:
\begin{equation}
\small{J=\begin{pmatrix}
-k_{-1}-k_2+2k_3X_0Y_0-3k_{-3}X_0^2~~~~ & k_{-2}+k_3X_0^2\\
k_2-2k_3X_0Y_0+3k_{-3}X_0^2~~~~ & -k_{-2}-k_3X_0^2
\end{pmatrix}.}
\end{equation}
Let $\sigma$ be the trace of $J$, given by $\sigma=\mathrm{tr}(J)=-(k_{-1}+k_2+k_{-2})+2k_3X_0Y_0-(3k_{-3}+k_3)X_0^2$, and $\Delta$ be the determinant of $J$, given by $\Delta=\det(J)=k_{-1}(k_{-2}+k_3X_0^2)$. The eigenvalues of $J$ read
\begin{equation}
\lambda=\frac{\sigma}{2}\pm\mathrm{i} \sqrt{\Delta-\frac{\sigma^2}{4}}
\end{equation}
At Hopf Bifurcation, $\sigma=0$, and the limit cycle oscillation requires that $\sigma>0$, $\Delta > \sigma^2/4$. The approximate frequency of the limit-cycle oscillation is the imaginary part of $\lambda$:
\begin{equation}
    \omega=\sqrt{\Delta-\frac{\sigma^2}{4}},
\end{equation}
and the approximate period is $P=\frac{2\pi}{\omega}$.
	
Since we are particularly interested in the period sensitivity of $k_2$, we choose $k_2$ as the control parameter. The bifurcation point $k_2^c$ that satisfies $\sigma(k_2=k_2^c)=0$ reads 
\begin{equation}
    \footnotesize{k_2^c=\left[ k_{-1}+k_{-2}+\left( 3k_{-3}+k_3-\frac{2k_3k_{-3}X_0^2}{k_{-2}+k_3X_0^2} \right)X_0^2\right]\frac{k_3X_0^2+k_{-2}}{k_3X_0^2-k_{-2}},}
\end{equation}
and the limit cycle oscillation requires that $k_2>k_2^c$. Thus, we get the explicit expression of the period,
\begin{equation}
P\approx \frac{2\pi}{\sqrt{k_{-1}(k_{-2}+k_3X_0^2)-\frac{1}{4}\left( \frac{k_3X_0^2-k_{-2}}{k_3X_0^2+k_{-2}}\right) ^2(k_2-k_2^c)^2}}.
\end{equation}
Then $C_2$, the period sensitivity of $k_2$, is given by
\begin{equation}
\begin{split}
C_2&\equiv\frac{\partial \ln P}{\partial\ln k_2}\\
&=\frac{k_2(k_2-k_2^c)}{4k_{-1}(k_{-2}+k_3X_0^2)\left( \frac{k_3X_0^2+k_{-2}}{k_3X_0^2-k_{-2}}\right)^2-(k_2-k_2^c)^2}.
\end{split}
\end{equation}
Obviously, $C_2$ is zero at the onset, increases with $k_2$, and decreases with $k_{-2}$. The leading order behavior of $C_2$ reads
\begin{equation}
C_2\simeq A(k_2-k_2^c),
\end{equation} 
where $A\equiv \frac{k_2^c(k_3X_0^2-k_{-2})^2}{4k_{-1}(k_{-2}+k_3X_0^2)^3}>0$.

\subsubsection{Far from the oscillation onset}

When $k_2$ is large, that is, $k_2\gg k_2^c$, as shown in Fig.~3 (d) of the main text, we consider the limit cycle on the $X$-$Z$ phase space, where $Z=X+Y$. Hence, Eqs. (\ref{eq:ODE_bruss}) can be written as 
\small{
\begin{subequations}{\label{eq:ODE_bruss_z}}
    \small{\begin{align}
        \frac{\mathrm{d} X}{\mathrm{d} t}&=k_1-(k_2+k_{-1})X+(k_{-2}+k_3X^2)(Z-X)-k_{-3}X^3,\\
        \frac{\mathrm{d} Z}{\mathrm{d} t}&=k_1-k_{-1}X. 
    \end{align}}
\end{subequations}}

Similar to the VdP model, the limit cycle of the Brusselator consists of two slow and two fast branches, which we refer to as S, S', and F, F' branches, as shown in Fig.~\ref{fig:S3}(b). We now estimate the scaling laws of the time spent on each branch. 

The two slow branches follow $X$-nullcline $\frac{\mathrm{d} X}{\mathrm{d} t}=0$, which reads
\begin{equation}
    Z=\frac{(k_2+k_{-1})X-k_1+k_{-3}X^3}{k_{-2}+k_3X^2}+X.\label{eq:X_null_bruss}
\end{equation}
The extrema of this nullcline satisfy $\frac{\mathrm{d} Z}{\mathrm{d} X}=0$, which are roots for the following polynomial,
\begin{equation}
    \epsilon X^4 + \left(3\frac{k_{-2}}{k_3}\epsilon -1\right) X^2 + \frac{2k_1}{k_3+k_{-3}}\epsilon X + \frac{k_{-2}}{k_3} = 0,\label{eq:poly_full}
\end{equation}
where $\epsilon = \frac{k_3+k_{-3}}{k_2+k_{-1}+k_{-2}}$. Since $\epsilon$ is small when $k_2$ is large and we are interested in the scaling behavior, we seek solutions of Eq.~(\ref{eq:poly_full}) in series of $\epsilon$ and focus on the leading term.

For simplicity, we first consider the irreversible limit where $k_{-2} = k_{-3} = 0$. The polynomial Eq.~(\ref{eq:poly_full}) becomes
\begin{equation}
    \epsilon X^3 - X + \frac{2k_1}{k_3}\epsilon = 0,\label{eq:poly_irre}
\end{equation}
where $\epsilon = \frac{k_3}{k_2+k_{-1}}\ll 1$. We expand $X$ in series as $X(\epsilon)=\epsilon^\rho\sum_{r=0}^{\infty}a_r\epsilon^r$ and substitute this expression into Eq.~(\ref{eq:poly_irre}), which yields two leading-order solutions with
\begin{equation}
    \rho = 1, a_0=2\frac{k_1}{k_3};\quad \rho=-\frac{1}{2}, a_0=1.
\end{equation}
Thus, the $X$-coordinates of the two extrema can be approximated by
\begin{equation}
    \small{X_1\approx 2\frac{k_1}{k_3}\epsilon = 2\frac{k_1}{k_2+k_{-1}}\sim \frac{k_1}{k_2}, \quad X_2\approx \epsilon^{-1/2} = \sqrt{\frac{k_2+k_{-1}}{k_3}} \sim \sqrt{\frac{k_2}{k_3}}.}
\end{equation}
Introducing these two values into Eq.~(\ref{eq:X_null_bruss}), we obtain the corresponding $Z$-coordinates of the two extrema,
\begin{equation}
    \small{Z_1 \approx \frac{(k_2+k_{-1})^2}{4k_1k_3} \sim \frac{k_2^2}{k_1k_3}, \quad Z_2 \approx 2\sqrt{\frac{k_2+k_{-1}}{k_3}}\sim \sqrt{\frac{k_2}{k_3}}.}
\end{equation}
Note that as $k_2$ increases, $Z_1$ grows faster than $Z_2$. Thus, the amplitude for the S-branch scales as
\begin{equation}
    \rm{amp}_\rm{s} \sim Z_1 - Z_2 \sim \frac{k_2^2}{k_1k_3}.
\end{equation}
The progression speed on the S-branch scales as
\begin{equation}
    v_\rm{s} \approx \abs{\frac{\mathrm{d} Z}{\mathrm{d} t}}\sim \abs{k_1-k_{-1}X_1} \sim k_1.
\end{equation}
Hence, the time spent on the S-branch scales as
\begin{equation}
    \tau_\rm{s} \sim \frac{\rm{amp}_\rm{s}}{v_\rm{s}} \sim \frac{k_2^2}{k_1^2k_3}.
\end{equation}

We can estimate the leading terms for the amplitude $\mathrm{amp}$, the progression speed $v$, and subsequently the duration $\tau$ of the other branches in a similar way. For the S'-branch, we have
\begin{equation}
    \begin{split}
        &\rm{amp}_\rm{s'} \sim Z_1\sim \frac{k_2^2}{k_1k_3},\\
        &v_\rm{s'} \sim \abs{k_1-k_{-1}X} \sim k_{-1}Z_1 \sim \frac{k_{-1}k_2^2}{k_1k_3}, 
    \end{split}
\end{equation}
which leads to
\begin{equation}
    \tau_\rm{s'} \sim \frac{\rm{amp}_\rm{s'}}{v_\rm{s'}} \sim \frac{1}{k_{-1}}.
\end{equation}
For the F-branch, we have
\begin{equation}
    \rm{amp}_\rm{f} \sim Z_1\sim \frac{k_2^2}{k_1k_3},\quad v_\rm{f} \sim \abs{\frac{\mathrm{d} X}{\mathrm{d} t}} \sim k_3Z_1^3 \sim \frac{k_2^6}{k_1^3k_3^2},
\end{equation}
resulting in
\begin{equation}
    \tau_\rm{f} \sim \frac{\rm{amp}_\rm{f}}{v_\rm{f}} \sim \frac{k_1^2k_3}{k_2^4}.
\end{equation}
Lastly, for the F'-branch, we have 
\begin{equation}
    \rm{amp}_\rm{f'} \sim X_2-X_1\sim \sqrt{\frac{k_2}{k_3}},\quad v_\rm{f'} \sim \abs{\frac{\mathrm{d} X}{\mathrm{d} t}} \sim k_2X_2 \sim \sqrt{\frac{k_2^3}{k_3}},
\end{equation}
and therefore, the time spent on the F'-branch scales as
\begin{equation}
    \tau_\rm{f'} \sim \frac{\rm{amp}_\rm{f'}}{v_\rm{f'}} \sim \frac{1}{k_2}.
\end{equation}

In the limit of large $k_2$, we obtain the order of four durations,
\begin{equation}
    \tau_\rm{s} \gg \tau_\rm{s'} \gg \tau_\rm{f'} \gg \tau_\rm{f}.\label{eq:time_order}
\end{equation}
We also note that $k_2$ indeed increases the progression speed of the S', F, and F-branches. However, their contribution to the period is decreased by the increase of $k_2$, thereby reducing their negative contribution to the period sensitivity of $k_2$. As a result, when far from the onset of oscillation, the period of an irreversible Brusselator can be approximated by
\begin{equation}
    P\approx \tau_\rm{s} \sim \frac{k_2^2}{k_1^2k_3},
\end{equation}
which gives the period sensitivities $C_2 \approx 2$, $C_1 \approx -2$, $C_3 \approx -1$, and $C_{-1} \approx 0$, consistent with the simulation results in Fig.~3(c) in the main text.

However, we notice that in the reversible Brusselator when $k_{-2}\neq 0$, the asymptotic behavior of $C_2$ is not 2 [Fig.~4(c) in the main text]. This is because another small number, $k_{-2}/k_3$, becomes non-negligible when $k_2$ is large enough such that $\epsilon \sim k_{-2}/k_3$ or even smaller. In this situation, we go back to Eq.~(\ref{eq:poly_full}) and seek solutions in the form $X(\epsilon)=\epsilon^\rho\sum_{r=0}^{\infty}a_r\epsilon^r$. What is different in this case is the non-homogeneous term $k_{-2}/k_3$ in Eq.~(\ref{eq:poly_full}). By focusing on the leading-order term, we find two solutions,
\begin{equation}
    \rho = 0, a_0=\sqrt{\frac{k_{-2}}{k_3}};\quad \rho=-\frac{1}{2}, a_0=1.
\end{equation}
The coordinates of the two extrema of the $X$-nullcline read
\begin{equation}
    \begin{split}
        (X_1,Z_1) &= \left(\sqrt{\frac{k_{-2}}{k_3}},~\frac{k_2}{2\sqrt{k_{-2}k_3}} \right),\\ (X_2,Z_2) &= \left(\sqrt{\frac{k_{2}}{k_3+k_{-3}}},~\sqrt{\frac{k_{2}}{k_3+k_{-3}}}(2+\frac{k_{-3}}{k_3}) \right).
    \end{split}
\end{equation}
In this reversible situation, as long as $k_{-2}$ is small enough to keep the system far from the onset, the four branches are well separated, and we can obtain the same order of the time intervals on each branch as Eq.~(\ref{eq:time_order}). The period can be estimated by the time spent on the S-branch as
\begin{equation}
    P\approx \tau_\rm{s} \sim \frac{Z_1 - Z_2}{\abs{\frac{\mathrm{d} Z}{\mathrm{d} t}}} \sim \frac{k_2}{k_1\sqrt{k_{-2}k_3}},
\end{equation}
which leads to $C_2\approx 1$, consistent with simulation results in Fig.~4(c) in the main text.

\subsection{Study of the Tyson Model for \textit{Drosophila} Circadian Clocks}

We study a simple model for \textit{Drosophila} circadian rhythms \cite{Tyson1999-lr}. This simplified version consists of two molecular species: the \textit{per} mRNA and the total PER protein. We denote their concentrations as $M$ and $P_\rm{T}$, respectively. The ODEs read \cite{Tyson1999-lr}
\begin{align}
    \frac{\rm{d}M}{\rm{d}t} &= \frac{v_\rm{m}}{1+\frac{P_2^2}{A^2}} - k_\rm{m}M,\label{eq:d_M}\\
    \frac{\rm{d}P_\rm{T}}{\rm{d}t} &= v_\rm{p} M - \frac{k_1P_1 + 2k_2P_2}{J + P_\rm{T}} - k_3P_\rm{T},\label{eq:d_PT}
\end{align}
where $P_1$ and $P_2$ represents monomers and dimers of PER protein, respectively, satisfying $P_1 = q P_\rm{T}$ and $P_2 = \frac{1-q}{2}P_\rm{T}$, where $q = \frac{2}{1+\sqrt{1+8KP_\rm{T}}}$, under the rapid-equilibrium assumption for their interconversion \cite{Tyson1999-lr}. $v_\rm{m},k_\rm{m},v_\rm{p},k_1,k_2$, and $k_3$ are kinetic rate constants, and we assume they are all temperature-dependent and obeying the Arrhenius law. $A$ and $J$ are Michaelis-Menten constants and presumably independent of temperature.

In this 2-species model, we can regard the protein $P_\rm{T}$ as the product in a substrate-depletion motif (like $X$ in Brusselator), and $M$ as the substrate (like $Y$ in Brusselator) \cite{Tyson2002-rf}. The nonlinear term in Eq.~\ref{eq:d_PT} with phosphorylation rates $k_1$ and $k_2$ involved is auto-catalysis of $P_\rm{T}$, and in some parameter regime, this auto-regulation can be auto-activation. On the other hand, $P_\rm{T}$ represses $M$ in a Michaelis-Menten form, while $M$ produces $P_\rm{T}$ linearly. 

In Fig.~\ref{fig:S5}(a), we demonstrate that TC can be achieved in this system by increasing $k_1$, the phosphorylation rate of PER monomer. This enhancement is due to the increase of the period sensitivity of $k_1$ that becomes relatively large and positive when it is far from the onset [Fig.~\ref{fig:S5}(b)]. However, unlike the saturation behavior in the VdP model or the Brusselator, the period sensitivies diverges at large $k_1$ because there is a singular Hopf bifurcation at around $\ln k_1 = 3.84$, where a small perturbation of $k_1$ results in a large variation of the amplitude [Fig.~\ref{fig:S5}(c)], a phenomenon known as ``Canard explosion" \cite{Eckhaus1983-zj}. Near this critical point, it is difficult to accurately compute period sensitivities numerically. 

The large positive period sensitivity of $k_1$ far from the onset is also a result from oscillation phase separation (OPT). As shown in Fig.~\ref{fig:S5}(d), the limit cycle can be separated into two phases: the S phase is when $P_\rm{T}$ is at a low level while $M$ increases along the $P_\rm{T}$-nullcline, and the F phase is the rest of the limit cycle when $P_\rm{T}$ varies rapidly. Fig.~\ref{fig:S5}(e) demonstrates that although the amplitudes of both phases scale with $k_1$ similarly, Only the F phase is accelerated by $k_1$. The $k_1$-increased amplitude with the $k_1$-decreased progression speed in the S phase is the key to the large positive period sensitivity of $k_1$.

Parameter values (at $T_0 = 298$ K) used in our simulations: $v_\rm{m} = 1$, $k_\rm{m} = 0.1$, $v_\rm{p} = 0.5$, $k_2 = 0.03$, $k_3 = 0.1$, $A=0.1$, $J =0.05$, and $K = 200$. Activation energies are 15 $k_\rm{B}T_0$ for all kinetic rate constants except for $k_1$, which is 25 $k_\rm{B}T_0$.

\subsection{Details of the VZH Model for Kai System}

The model was originally introduced by Hatakeyama \& Kaneko to investigate TC in the Kai system \cite{Hatakeyama2012-pq}.
Here, we slightly modify the original model to include all the kinetic rates from elementary reactions [Fig.~5(a) in the main text], and derive the following ODEs:
\small{
\begin{equation}
    \frac{\rm{d}[C_i]}{\rm{d}t} = k_{\rm{u}i} [AC_i] - k_{\rm{b}i} [A][C_i] +\delta_{i,0} b_0 [\Tilde{C}_i] - \delta_{i,6} f_6 [C_i],
\end{equation}
\begin{equation}
    \begin{split}
        \frac{\rm{d}[AC_i]}{\rm{d}t} &= (1-\delta_{i,0})k_{\rm{p}i} [AC_{i-1}] - (1-\delta_{i,6})k_{\rm{p},i+1} [AC_{i}] \\
        &\quad  - k_{\rm{u}i} [AC_i] + k_{\rm{b}i} [A][C_i],
    \end{split}
\end{equation}
\begin{equation}
    \begin{split}
        \frac{\rm{d}[\Tilde{C}_i]}{\rm{d}t} &= -(1-\delta_{i,0})k_{\rm{dp}i} [\Tilde{C}_{i}] + (1-\delta_{i,6})k_{\rm{dp},i+1} [\Tilde{C}_{i+1}] \\
        &\quad -\delta_{i,0} b_0 [\Tilde{C}_i] + \delta_{i,6} f_6 [C_i],
    \end{split}
\end{equation}}
where $\delta_{ij}$ is the Kronecker delta function, and the index $i$ runs from 0 to 6. The free KaiA concentration $[A]$ is given by the constraint:
\begin{equation}
    [A] + \sum_{i=0}^6[AC_i] = [KaiA]_\rm{total}.
\end{equation}
For the model parameters, we adopted the same differential binding affinity strategy as Hatakeyama \& Kaneko \cite{Hatakeyama2012-pq}, namely, $k_{\rm{u}i} = k_{\rm{u}0} \alpha^i$, and $k_{\rm{b}i}=k_{\rm{b}0}$, from $i=0$ to 6. For results in Fig.~5 in the main text, the exact parameters (at $T=T_0 = 298$K) are given by
\begin{align*}
    k_{\rm{p}} &= [1.03,1.03,1.03,1.03,1.44,1.24]~\rm{s}^{-1},\\
    k_{\rm{dp}} &= [3.16,3.16,3.16,3.16,3.16,3.16]~\rm{s}^{-1},\\
    k_{\rm{u}0} &= 3.33~\rm{s}^{-1},\quad \alpha = 10,\quad  k_{\rm{b}0} = 1.66\times 10^7~\rm{nM}^{-1}\rm{s}^{-1},\\
    b_0 &= f_6 = 0.333~\rm{s}^{-1},\quad [C]_\rm{total} = 11.4~\rm{nM},
\end{align*}
and we have verified that the qualitative behaviors of $Q_{10}$ and period sensitivities are robust in a wide range of parameters. The slightly modified values in $k_\rm{p}$ is only for fitting the curve with the experimental data in Fig.~5(b) in the main text.

\newpage
\section{Supplementary Figures}

\begin{figure}[!h]
\centering
\includegraphics[width=1\textwidth]{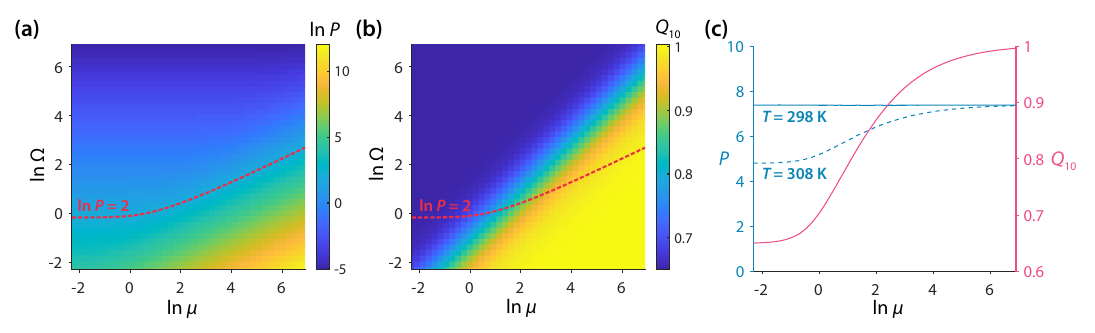}
\caption{\label{fig:S1} Enhancing TC under a constant period in the VdP model. (a) The period ($P$) at the room temperature $T_0=298$ K is determined by the two kinetic rates $\mu$ and $\Omega$. (b) With a fixed choice of activation energies, the $Q_{10}$ value of the period depends on both $\mu$ and $\Omega$. Here, $E_\mu = 26.7$ and $E_\Omega = 13.3$  ($k_\rm{B} T_0$). (c) Along the contour line of the period ($\ln P =2$) in (a), the period at $T=308$ K approaches that at $T=298$ K with increasing $\ln \mu$, leading to an increase of $Q_{10}$ toward 1.} 
\end{figure}

\begin{figure}[!h]
\centering
\includegraphics[width=0.7\textwidth]{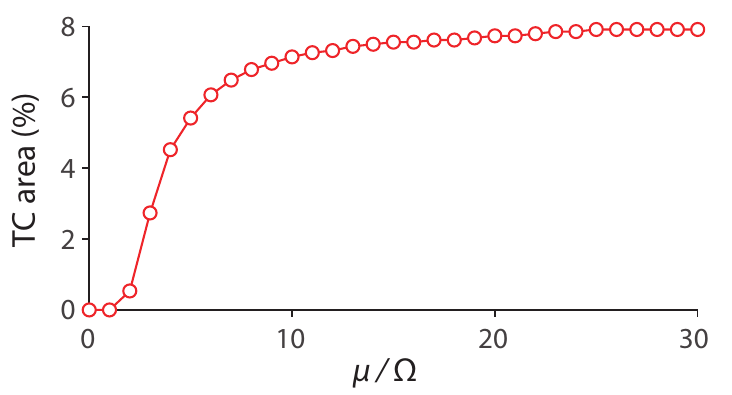}
\caption{\label{fig:S2} In the VdP model, the area of the TC regime in the activation-energy space [highlighted in Fig.~1(c) in the main text] varies with the ratio $\mu/\Omega$ at $T=T_0$.} 
\end{figure}

\begin{figure}
\centering
\includegraphics[width=0.7\textwidth]{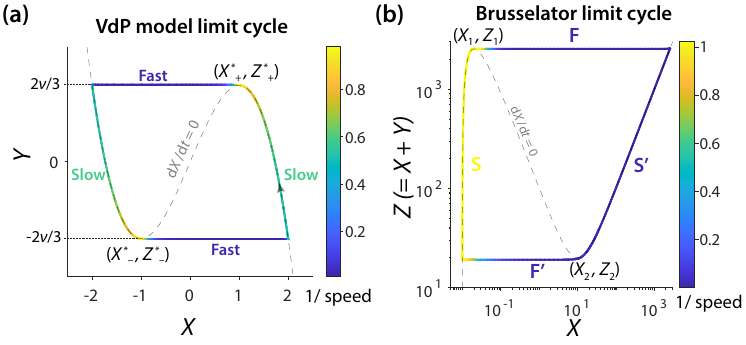}
\caption{\label{fig:S3} Limit cycles in the VdP model and the irreversible Brusselator far from the onset. (a) The VdP model limit cycle at $\nu=\mu/\Omega \gg 1$. The orbit exhibits four branches: two identical slow branches(with opposite signs in the progression velocity) and two identical fast branches. The X-nullcline extrema $(-1,-\frac{2}{3}\nu)$ and $(1,\frac{2}{3}\nu)$ determine the amplitude. (b) The Brusselator limit cycle at $k_2\gg k_2^c$ where $k_2^c$ is the Hopf bifurcation point. Similar to the VdP model, the orbit is also separated into four branches: S, S', F, and F'. The X-nullcline extrema $(X_1,Z_1)$ and $(X_2,Z_2)$ determines the amplitude of the branches. } 
\end{figure}

\begin{figure}
\centering
\includegraphics[width=1\textwidth]{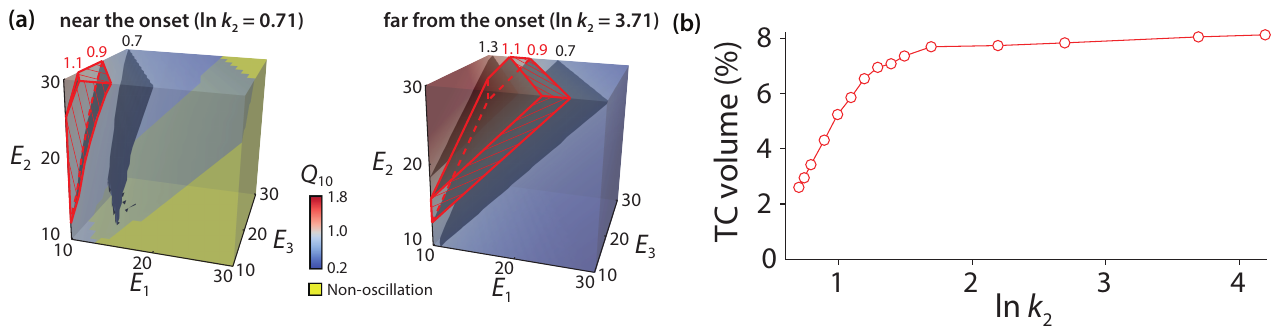}
\caption{\label{fig:S4} TC is enhanced by the increase of $k_2$ under different activation energy settings in the irreversible Brusselator. (a) 
The $Q_{10}$ values of the period near the onset (left) and far from the onset (right) at varying values of activation energies $E_1$, $E_2$, and $E_3$ (in the unit of $k_\rm{B}T_0$).
$E_{-1}$ is fixed at 20~$k_\rm{B}T_0$. The highlighted regime marked in red is the temperature-compensated regime. (b) The ``TC volume", defined by the volume of the temperature-compensated regime in the $E_1$-$E_2$-$E_3$ parameter space highlighted in (a), increases with $k_2$. } 
\end{figure}

\begin{figure}
\centering
\includegraphics[width=0.7\textwidth]{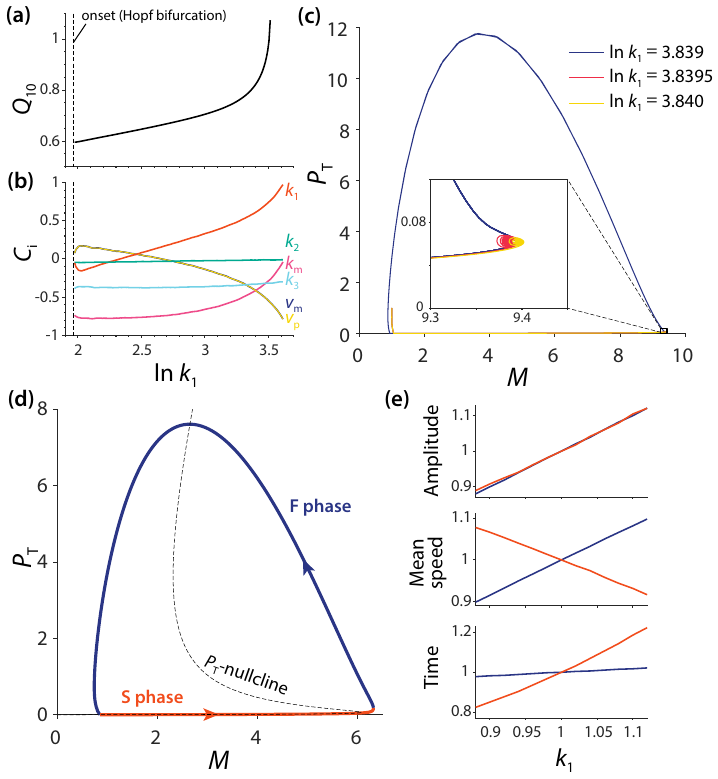}
\caption{\label{fig:S5} TC and oscillation phase separation in the Tyson model for \textit{Drosophila} circadian clock. (a) $Q_{10}$ varies with $k_1$. (b) Period sensitivities varies with $k_1$. Note that the period sensitivity of $k_1$ becomes relatively large and positive compared to other period sensitivities. (c) Trajectories near another Hopf bifurcation point at large $k_1$. The divergent behavior of $Q_{10}$ and $C_i$'s at large $k_1$ comes from the singularity of the Hopf bifurcation point around $\ln k_1 = 3.84$, where a small perturbation of $k_1$ results in a large variation of the amplitude, a phenomenon known as ``Canard explosion" \cite{Eckhaus1983-zj}. (d) The limit cycle when $ \ln k_1 = 3.4$ (far from the onset) is well separated into two phases: the S phase where $P_\rm{T}$ keeps low and $M$ increases, and the F phase where $P_\rm{T}$ varies dramatically. As a practical definition, we define S phase as from the minimum of $P_\rm{T}$ to the maximum of $M$. (e) The amplitude, mean progression speed, and the time on each of the two phases varies with $k_1$ differently.
} 
\end{figure}

\begin{figure}
\centering
\includegraphics[width=1.0\textwidth]{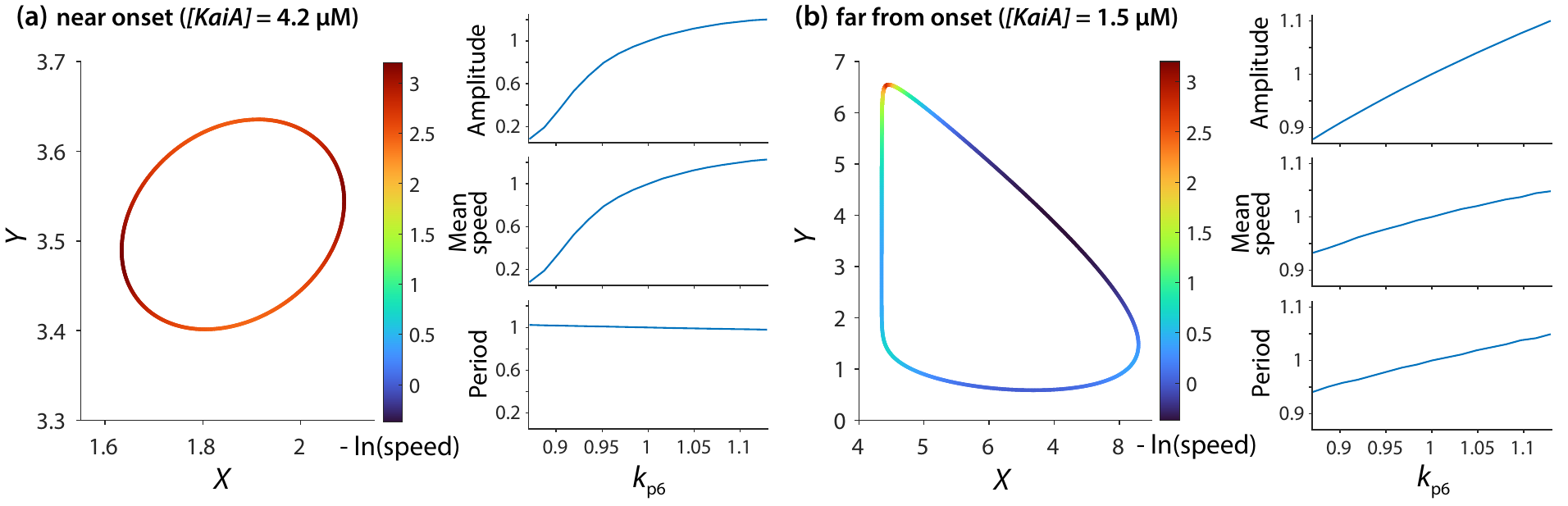}
\caption{\label{fig:S6} The limit cycle properties near Hopf bifurcation and far from Hopf bifurcation in the vZH model. (a) Near Hopf bifurcation ($[KaiA]_\rm{total}=4.2~\mu$M). Left: the limit-cycle trajectory of KaiC oscillation projected on the X-Y subplane where $X= [C_5]+[AC_5]$ and $Y = \sum_{i=0}^4 [C_i]+[AC_i]$. The color shows the inverse of the progression speed in the log scale. Right: the amplitude (perimeter of the limit cycle), mean progression speed, and the total period varies with $k_\rm{p6}$. Both the amplitude and the progression speed increases with $k_\rm{p6}$ in a similar manner, resulting in a period insensitive to $k_\rm{p6}$. (b) Far from Hopf bifurcation ($[KaiA]_\rm{total}=1.5~\mu$M). The mean progression speed scales with $k_\rm{p6}$ more slowly than the amplitude (due to different scaling behavior in different phases shown in Fig.~5 in the main text), resulting in a period increased by $k_\rm{p6}$.} 
\end{figure}

\end{document}